\documentstyle[12pt] {article}
\newcommand{\beq}{\begin{equation}}
\newcommand{\eeq}{\end{equation}}
\newcommand{\beqa}{\begin{eqnarray}}
\newcommand{\eeqa}{\end{eqnarray}}
\newcommand{\ba}{\begin{array}}
\newcommand{\ea}{\end{array}}

\begin{document}
  
\begin{center}
{\large \bf Quantum Transition from Order to Chaos\\
in the Nuclear Shell Model}
\footnote{To be published in the Proceedings of the International 
Conference "{\it Chaos, Fractals and Models '96}", University 
of Pavia (Italy), October 25--27, 1996.} 
\end{center}

\vskip 0.5 truecm

\begin{center}
{\bf Luca Salasnich}\footnote{E-mail: salasnich@math.unipd.it}
\vskip 0.5 truecm
Dipartimento di Matematica Pura ed Applicata, Universit\`a di Padova, \\
Via Belzoni 7, I--35131 Padova, Italy \\
Istituto Nazionale di Fisica Nucleare, Sezione di Padova, \\
Via Marzolo 8, I--35131 Padova, Italy \\
Istituto Nazionale di Fisica della Materia, Unit\`a di Milano, \\
Via Celoria 16, 20133 Milano, Italy
\end{center}

\vskip 1. truecm

{\bf Abstract.} We discuss the role of quantum chaos in atomic nuclei. 
After reviewing the basic assumptions of the nuclear shell 
model, we analyze the spectral statistics of the energy levels 
obtained with realistic shell--model calculations in the $fp$ shell. 
In particular, for Ca isotopes we observe a transition from order to 
chaos by increasing the excitation energy and a clear quantum 
signature of the breaking of integrability by changing 
the single--particle spacings. 

\vskip 1. truecm

\section{Chaos and Quantum Chaos}
\par
Quantum chaos is the study of the properties of 
quantal systems which are classically chaotic, thus 
with exponential divergence of initially closed trajectories 
in the phase space [1]. 
\par
The energy fluctuation properties of systems with underlying 
classical chaotic behaviour and time--reversal symmetry agree with 
the predictions of the Gaussian Orthogonal Ensemble (GOE) of 
random matrix theory, whereas quantum analogs of classically 
integrable systems display the characteristics 
of the Poisson statistics [2]. 
\par
The most used quantity to study the local fluctuations of the energy levels 
is the spectral statistics $P(s)$. $P(s)$ is 
the distribution of nearest--neighbour spacings 
$s_i=({\tilde E}_{i+1}-{\tilde E}_i)$ of the unfolded levels ${\tilde E}_i$. 
It is obtained by accumulating the number of spacings that lie within 
the bin $(s,s+\Delta s)$ and then normalizing $P(s)$ to unity. 
For quantum systems whose classical analogs are integrable, 
$P(s)$ is expected to follow the Poisson limit, i.e. 
$P(s)=\exp{(-s)}$. On the other hand, 
quantal analogs of chaotic systems exhibit the spectral properties of 
GOE with $P(s)= (\pi / 2) s \exp{(-{\pi \over 4}s^2)}$, which is the 
so--called Wigner distribution. 
The distribution $P(s)$ is the best spectral statistics to analyze 
shorter series of energy levels and 
intermediate regions between order and chaos [1,2]. 

\section{The Nuclear Shell Model}
\par
One of the best systems for the study of quantum chaos is the atomic 
nucleus. In fact, its experimental energy levels 
have been studied in the domain of neutron and proton resonances 
near the nucleon emission threshold, 
where a large number of levels with the same spin and 
parity in the same nucleus are present, and an excellent agreement with GOE 
predictions has been found [3]. 
\par
In the nuclear shell model [4], the nuclear states, 
like the electronic states in atoms, are 
described in terms of the motion of nucleons in a mean--field. 
But, while the nuclear field is generated by the interactions 
of the nucleons (protons and neutrons), the atomic field is mainly 
governed by the interaction of the electrons with the nucleus. 
\par
The Hamiltonian of the shell model can be written as:
\beq
H_0=\sum_{i=1}^A \Big( -{\hbar^2\over 2m_i} \nabla_i^2 
+U_i \Big) \; ,
\eeq
where A is the number of nucleons, $m_i$ is the mass of the 
$i$-th nucleon, and $U_i$ is the
mean--field. The choice of the mean--field is crucial; $U_i$ may 
be obtained by the usual methods using the many body theory [4]. 
\par
To obtain a good agreement between the shell model results and the 
experimental data, it is necessary to add a residual interaction 
$H_R$ so that the total hamiltonian $H$ can be written: 
\beq
H=H_0+H_R \; , 
\eeq
where $H_R$ is the part of nucleon--nucleon interaction 
not included in $H_0$. Using second quantization formalism [4], 
we can write:
\beq
H_0=\sum_{\alpha} \epsilon_{\alpha}a_{\alpha}^+a_{\alpha} \; ,
\eeq
\beq
H_R=\sum_{\alpha \beta \gamma \delta} <\alpha \beta|V|\delta \gamma >
a_{\alpha}^+ a_{\beta}^+ a_{\gamma} a_{\delta} \; ,
\eeq 
where the labels denote the accessible single--particle states, 
$\epsilon_{\alpha}$ is the corresponding single--particle energy, 
and $<\alpha \beta|V|\delta \gamma >$ is the two--body matrix element 
of the nuclear residual interaction. The operators 
$a_{\alpha}^{+}$ and $a_{\alpha}$ are the fermionic creation and
annihilation operators of the $\alpha$--th single nucleon state and 
$H_R$ can be calculated by the 
Hartree--Fock equations, starting from the free nucleon interactions. 
Many nucleons are frozen in the deeper shells of the mean field potential 
and form an inert core; only a few nucleons partially populate the single
particle shells outside the core. These are called valence--nucleons. 
So there are N valence--nucleons, m active shells and a finite
number of energy levels. It is standard procedure to cut the basis 
states by introducing a finite number of configurations which are 
sufficient to describe the first excitation states [4]. 

\section{Spectral Statistics and the Brody Distribution}
\par
We study the ($f_{7/2}$,$p_{3/2}$,$f_{5/2}$,$p_{1/2}$) 
shell--model space, assuming $^{40}$Ca as an inert core with 
$4<N<10$ valence--nucleons. 
The diagonalizations are performed by using 
Lanczos and Householder algorithms with the code ANTOINE [5,6]. 
For a fixed number of valence protons and neutrons 
we calculate the energy spectrum for 
total angular momentum $J$ and total isospin $T$. 
The single--particle energies (in MeV) are $\epsilon_{7/2}=0$, 
$\epsilon_{3/2}=2$, $\epsilon_{1/2}=4$ and $\epsilon_{5/2}=6.5$. 
The residual interaction we use is a minimally modified Kuo--Brown 
realistic force with monopole improvements [6]. 
\par
For the low--lying levels, the 
spectrum is mapped onto unfolded levels with 
quasi--uniform level density by using both 
the constant temperature formula [7] and the local unfolding method [8]. 
The two procedures give the same results. 
For the full spectrum we use the local unfolding 
because the constant temperature formula is valid only 
when the level density grows exponentially. 
\par
To quantify the chaoticity of the distribution $P(s)$ (nearest--neighbour 
spacings of the energy levels) in terms of a parameter, 
we compare it to the Brody distribution
\beq
P(s,\omega)=\alpha (\omega +1) s^{\omega} \exp{(-\alpha s^{\omega+1})} \; ,
\eeq
with 
\beq
\alpha = (\Gamma [{\omega +2\over \omega+1}])^{\omega +1} \; . 
\eeq
This distribution interpolates between the Poisson distribution ($\omega =0$) 
of regular systems and the Wigner distribution ($\omega =1$) 
of chaotic ones (GOE). The parameter $\omega$ can be used as a simple 
quantitative measure of the degree of chaoticity [9]. 
\par  

\begin{center}
{\bf Table 1.} Brody parameter $\omega$ for Ca isotopes. 
\vskip 0.3 truecm
\begin{tabular}{|ccccccc|} \hline\hline 
$^{44}$Ca & $^{45}$Ca & $^{46}$Ca & $^{47}$Ca & $^{48}$Ca 
& $^{49}$Ca & $^{50}$Ca \\ 
\hline
0.69 & 0.75 & 0.99 & 0.98 & 0.95 & 1.00 & 0.87 \\ 
\hline\hline
\end{tabular}
\end{center}

Table 1 shows the Brody parameter $\omega$ for the whole spectrum 
of the analyzed Ca isoptopes (only neutrons in the 
$fp$ shell), which range from $^{44}Ca$ to $^{50}Ca$.  
We see that only the lightest Ca isotopes are not fully chaotic.  
\par
It now becomes interesting to analyze the Brody parameter as a function 
of the excitation energy. 
We calculate the $P(s)$ distribution and the Brody parameter 
up to a fixed value of the excitation energy above the yrast lines. 
Obviously this cannot be done 
for the lightest Ca isotopes because few levels are involved. 

\begin{center}
{\bf Table 2.} Brody parameter $\omega$ as a function the the 
Energy for Ca isotopes.  
\vskip 0.3 truecm
\begin{tabular}{|cccc|} \hline\hline 
Energy (MeV) & $^{48}$Ca & $^{49}$Ca & $^{50}$Ca\\ 
\hline
 6 & 0.54 & 0.63 & 0.72   \\ 
 8 & 0.65 & 0.70 & 0.79   \\ 
10 & 0.78 & 0.81 & 0.84   \\ 
12 & 0.83 & 0.82 & 0.86   \\ 
14 & 0.90 & 0.86 & 0.88   \\ 
16 & 0.87 & 0.92 & 0.90   \\
18 & 0.93 & 0.95 & 0.92   \\ 
\hline\hline
\end{tabular}
\end{center}

The results are written in Table 2 for $^{48}Ca$, $^{49}Ca$ and $^{50}Ca$ 
and show a strong energy dependence: there is an order--chaos 
transition by increasing the excitation energy. 
\par 
Another important aspect is the effect of the one--body Hamiltonian on 
the two--body residual interaction when the single--particle spacings are 
changed. 

\begin{center}
{\bf Table 3.} Brody parameter $\omega$ for Ca isotopes. 
\vskip 0.3 truecm
\begin{tabular}{|ccc|} \hline\hline 
Single Particle Spacings & $^{44}$Ca & $^{45}$Ca \\ 
\hline
degenerate        & 0.85 & 0.98  \\ 
normal            & 0.69 & 0.75  \\ 
double--spaced    & 0.29 & 0.58  \\ 
\hline\hline
\end{tabular}
\end{center}

Table 3 shows that with degenerate single--particle levels 
the isotopes $^{44}Ca$ and $^{45}Ca$ are chaotic while 
for double--spaced single--particle 
levels they are quasi--regular. This interesting effect 
is a clear quantum signature of the breaking of the integrability 
due to the residual interaction. In fact, the one--body Hamiltonian 
is classically integrable because it is the sum of harmonic oscillators 
while the two--body residual interaction is strongly non--linear. 
By increasing the single--particle spacings, the single--particle 
mean--field motion in the valence orbits suffers less disturbance and 
is thus more regular. 

\section{Summary}
\par
We have studied quantum chaos in atomic nuclei by using 
the nuclear shell model. We have calculated 
the spectral statistics of Ca isotopes 
by using Lanczos and Householder algorithms with the shell--model code ANTOINE. 
The Brody parameter, which fits the nearest--neighbour 
spacings distribution of the energy levels, shows an order--chaos transition 
when the excitation energy is increased and when the single--particle 
spacings are changed. 

\section*{References}

\parindent=0. pt

[1] M.C. Gutzwiller, {\it Chaos in Classical and Quantum Mechanics} 
(Springer--Verlag, Berlin, 1990).

[2] O. Bohigas and H.A. Weidenm\"uller, Ann. Rev. Nucl. Part. Sci. 
{\bf 38} (1988) 421 

[3] M.T. Lopez--Arias, V.R. Manfredi and L. Salasnich, 
Riv. Nuovo Cimento {\bf 17}, N. 5 (1994) 1.

[3] W.E. Ormand and R.A. Broglia, Phys. Rev. C {\bf 46} (1992) 1710; 
V. Zelevinsky, M. Horoi and B. A. Brown, Phys. Lett. {\bf 350} (1995) 141; 
M. Horoi, V. Zelevinsky, B. A. Brown, Phys. Rev. Lett. {\bf 74} (1995) 5194.

[4] R.D. Lawson, {\it Theory of the Nuclear Shell Model} 
(Clarendon, Oxford 1980).

[5] E. Caurier, computer code ANTOINE, C.R.N., Strasbourg (1989); 
E. Caurier, A. P. Zuker and A. Poves: in {\it Nuclear Structure of 
Light Nuclei far from Stability: Experiment ad Theory}, Proceedings of 
the Obernai Workshop 1989, Ed. G. Klotz (C.R.N, Strasbourg, 1989).

[6] E. Caurier, J.M.G. Gomez, V.R. Manfredi and L. Salasnich, 
Phys. Lett. B {\bf 365} (1996) 7.

[7] J.F.Jr. Shriner, G.E. Mitchell and T. von Egidy, 
Z. Phys. {\bf 338} (1991) 309.

[8] V.R. Manfredi, Lett. Nuovo Cimento {\bf 40} (1984) 135.

[9] T.A. Brody, Lett. Nuovo Cimento {\bf 7} (1973) 482. 

\end{document}